\documentclass{cjaa}                   

\input{epsf.sty}                        
\input{psfig.sty}                       

\begin{document}

   \title{A Detailed Study on the Equal Arrival Time Surface Effect in Gamma-Ray Burst Afterglows
}

   \volnopage{Vol.0 (200x) No.0, 000--000}      
   \setcounter{page}{1}          

   \author{Yong-Feng HUANG
      \inst{1,2}\mailto{}
   \and Ye LU
      \inst{3,2}
   \and Anna Yuen Lam WONG 
      \inst{2}
   \and Kwong Sang CHENG
      \inst{2}
      }
   \offprints{Y.-F. Huang}                   

   \institute{Department of Astronomy, Nanjing University, Nanjing 210093, China\\
             \email{hyf@nju.edu.cn}
        \and
             Department of Physics, The University of Hong Kong, Pokfulam Road, Hong Kong, China\\
        \and
             National Astronomical Observatories, Chinese Academy of Sciences, Beijing 100012, China\\
          }

   \date{Received~~2006 month day; accepted~~2006~~month day}

   \abstract{
Due to the relativistic motion of gamma-ray burst remnant and its 
deceleration in the circumburst medium, the equal arrival time surfaces 
at any moment are not spherical, but should be distorted ellipsoids. 
This will leave some imprints in the afterglows. In this article, we 
study the effect of equal arrival time surfaces numerically under 
various conditions, i.e., for isotropic fireballs, collimated jets, 
density jump conditions, and energy injection events. For each condition,
direct comparison between the two instances when the effect is and is 
not included, is presented. For isotropic fireballs and jets viewed 
on axis, the effect slightly hardens the spectra and postpones the peak 
time of afterglows, but does not change the shapes of the spectra and 
light curves significantly. In the cases when a density jump or an 
energy injection is involved, the effect smears the variability of 
the afterglows markedly.  
   }

   \authorrunning{Y.-F. Huang et al.}            
   \titlerunning{Equal arrival time surfaces in GRB afterglows }  

   \maketitle

%
%
\section{Introduction}

Afterglow observations have made it clear that gamma-ray bursts (GRBs), 
both long and short, typically lie at cosmological distances (Costa 
et al. 1997; Frail et al. 1997; Galama et al. 1997; Vreeswijk et al. 1999; 
Hjorth et al. 2002; Villasenor et al. 2005; Fox et al. 2005), with the 
highest redshift recorded so far being $z \sim 6.3$ for GRB 050904 
(Tagliaferri et al. 2005; Haislip et al. 2006; Price et al. 2006; 
Watson et al. 2006; Cusumano et al. 2006). Evidence is also accumulating, 
supporting the idea that long/soft GRBs may come from the collapse of 
massive stars, while short/hard GRBs come from the merger of two 
compact objects (Barthelmy et al. 2005). As the most violent bursts in 
the Universe since the Big Bang, GRBs and their afterglows can be 
satisfactorily understood in the framework of the relativistic fireball
model, which postulates that the main burst emission should  
be due to internal shocks and the afterglow emission can be accounted for 
by external shocks (M\'esz\'aros \& Rees 1992, 1997; Sari, Narayan, \& 
Piran 1996; Vietri 1997; Wijers, Rees, \& M\'esz\'aros 1997; Sari, Piran, 
\& Narayan 1998; Dermer, Chiang, \& B\"ottcher 1999; Su et al. 2006; and 
for recent reviews, see: van Paradijs, Kouveliotou, \& Wijers 2000; 
Piran 2004; Zhang \& M\'esz\'aros 2004). 

GRBs are one of the most relativistic phenomena in our cosmos. The initial 
bulk Lorentz factor of GRB ejecta can be as high as 100 --- 1000. Such 
an ultra-relativistic motion imposes two effects on the afterglows. First, 
the emission is strongly enhanced toward the direction of motion due to 
relativistic boosting. Second, photons emitted simultaneously from a 
spherical surface of the GRB remnant do not reach the observer at the same
time. Photons at higher latitude will arrive later. In other words, at 
any lab-frame time, while the shape of the GRB remnant itself is 
spherical, the photons received by the observer actually do not come 
from a spherical surface, but from a distorted ellipsoid, i.e., the 
equal arrival time surface (Waxman 1997; Panaitescu \& M\'esz\'aros 
1998; Sari 1998; Granot, Piran, \& Sari 1999; Gao \& Huang 2006). 
Additionally, if angularly resolved by a telescope, the equal arrival 
time surface (EATS) would not be homogeneous in brightness, but would 
show a ring-like structure. 

The exact analytical expressions for the geometric shape of EATS can 
be derived under some simplified assumptions in the ultra-relativistic stage, 
for example, in the cases of fully radiative and adiabatic regimes 
(Bianco \& Ruffini 2005). But the EATS effect on the emission can be 
well incorporated in modeling of GRB afterglows only through numerical 
calculations. This has been done by a few authors (Panaitescu \& 
M\'esz\'aros 1999; Moderski, Sikora, \& Bulik 2000; Huang et al. 2000a, 
2000b; Salmonson 2003; Kumar \& Granot 2003; Granot 2005). However, 
a direct comparison between the two instances where the EATS influence is 
and is not included, which can reveal the effect more apparently, is still
lacking. In this article, we intend to carry out the comparison. The 
structure of our paper is organized as follows. We describe our model in 
\S 2. Numerical results are then presented in \S 3 under various 
conditions, for example, for isotropic fireballs, jets, energy 
injections, and density variations in the circumburst medium. Finally,
\S 4 is our conclusion and discussion.

\section{Model Description}

According to the standard fireball theory, afterglows are produced when
the GRB ejecta, either isotropic or highly collimated, ploughs through
the circumburst medium, producing a strong blastwave that accelerates 
the swept-up electrons. Synchrotron emission from these electrons is the 
dominant radiation mechanism that takes effect in the afterglow 
stage, although inverse Compton scattering may also play a role in some 
cases (Wei \& Lu 2000; Sari \& Esin 2001). The GRB ejecta is initially 
ultra-relativistic, but may become trans-relativistic in a few months 
(Huang et al. 1998), and enter the deep Newtonian phase in two or three
years (Huang \& Cheng 2003). Additionally, the blastwave is in the 
highly radiative regime in the initial few hours, but will be 
adiabatic thereafter. 

A simple model that can realistically depict the overall evolution of 
GRB afterglows and which is also very convenient to solve numerically 
has been proposed by Huang et al. (Huang et al. 1999, 2000a, 2000b; 
Huang \& Cheng 2003). We will use this model for the current study. 
Now we first describe the model briefly for completeness. In the 
description below, unless declared explicitly, physical quantities 
are all measured in the observer's static lab frame. 

The model 
is mainly characterized by a generic dynamical equation of 
(Huang et al. 1999), 
\begin{equation}
\label{dgdm1}
\frac{d \gamma}{d m} = - \frac{\gamma^2 - 1}
       {M_{\rm ej} + \epsilon m + 2 ( 1 - \epsilon) \gamma m}, 
\end{equation}
where $\gamma$ is the bulk Lorentz factor of the shocked medium, 
$m$ is the swept-up mass, $M_{\rm ej}$ is the initial 
mass of the GRB ejecta, and $\epsilon$ is the radiative efficiency. 
Equation~(1) is applicable in both the
ultra-relativistic and the non-relativistic phases (Huang et al. 1999). 
For collimated GRB ejecta, the lateral expansion is 
described realistically by (Huang et al. 2000a, 2000b),
\begin{equation}
\label{dthdt2}
\frac{d \theta}{d t} = \frac{c_{\rm s} (\gamma + \sqrt{\gamma^2 - 1})}{R},
\end{equation}
where $\theta$ is the half-opening angle of the jet, $R$ is the 
radius of the shock, and $t$ is observer's time. 
$c_{\rm s}$ is the comoving sound speed, which is
further given by
\begin{equation}
\label{cs3}
c_{\rm s}^2 = \hat{\gamma} (\hat{\gamma} - 1) (\gamma - 1) 
	      \frac{1}{1 + \hat{\gamma}(\gamma - 1)} c^2 , 
\end{equation}
with $\hat{\gamma} \approx (4 \gamma + 1)/(3 \gamma)$ being the adiabatic 
index.

To calculate synchrotron radiation from the shock-accelerated electrons, 
the electron distribution function is a key factor. Basically the 
electrons follow a power-law distribution according to their energies, 
with the power-law index $p$ typically varying between 2 and 3. Here we adopt a 
refined function that takes into account the cooling effect 
(Dai, Huang, \& Lu 1999; Huang \& Cheng 2003).
Note that our distribution function is applicable even in the 
deep Newtonian phase (Huang \& Cheng 2003). As usual, we assume that 
the energy ratios of electrons and magnetic field with respect to 
protons are $\xi_{\rm e}$ and $\xi_{\rm B}$ respectively. 

In order to include the EATS effect, the observed afterglow flux 
density at any given time $t$ should be calculated by integrating 
over the EATS determined by 
\begin{equation}
\label{eqt28}
\int \frac{1 - \beta \cos \Theta}{\beta c} dR \equiv t,
\end{equation}
within the ejecta boundaries (Moderski et al. 2000), where 
$\beta = \sqrt{\gamma^2 -1}/ \gamma$ and  $\Theta$ is the latitude 
angle on the EATS. In our model, it is also very convenient to remove
the consideration on EATS, so that we can clearly see how the EATS
takes effect in GRB afterglows. For details on how to calculate the 
dynamics and the radiation process, readers may refer to Huang et al.'s
original articles (Huang et al. 1999, 2000a, 2000b; Huang \& Cheng 2003).

\section{Numerical Results}

In this section, we use our model to investigate the EATS effect in 
GRB afterglows under various conditions. In each condition, 
we will directly compare the 
two instances where the EATS influence is and is not included. For 
convenience, we first define a set of ``standard'' parameters that will 
be generally used in our calculations: $\xi_{\rm e}=0.1, \xi_{\rm B} =
0.001, p=2.5$, the isotropic equivalent energy of the GRB ejecta 
$E_{\rm 0,iso} = 10^{53}$ ergs, the initial Lorentz factor 
$\gamma_0 = 300$, the number density of the circumburst medium
$n=1$ cm$^{-3}$, and the luminosity distance of the GRB $D_{\rm L} =1$ 
Gpc. For jets, we take the initial half-opening angle as $\theta_0 =0.1$. 
These parameter values are quite typical in GRB afterglows. 

\subsection{Isotropic Fireballs}

In Figure 1, we illustrate the evolution of the afterglow spectrum for 
an isotropic fireball with ``standard'' parameters. The solid lines 
correspond to the instance when the EATS effect is included, while the 
dashed lines correspond to the case when the EATS effect is omitted. 
A few interesting features can be clearly seen from this figure. First, 
the spectrum at any particular moment can be divided into three 
segments. Taking the spectrum at $t=10^5$ s as an example, the three 
segments are approximately $S_{\nu} \propto \nu^{0.34}, \nu^{-0.76},
\nu^{-1.27}$ respectively. They are in good agreement with theoretical 
expectations, i.e., $S_{\nu} \propto \nu^{1/3}, \nu^{(1-p)/2}, \nu^{-p/2}$
(Sari, Piran, \& Narayan 1998). Note that the EATS does not change the 
slope of each segment. Secondly, the peak flux density ($S_{\nu,{\rm max}}$) does 
not evolve significantly with time. This is true irrespective of the EATS consideration. 
However, the inclusion of the EATS does reduce $S_{\nu,{\rm max}}$ by a factor 
of $\sim 2$. Thirdly, the EATS effect makes the spectrum slightly harder. 
As a result, the peak frequency $\nu_{\rm max}$ (corresponding to 
$S_{\nu,{\rm max}}$) is slightly higher, and the emission below $\nu_{\rm max}$
is reduced while that above $\nu_{\rm max}$ is enhanced. This effect is 
easy to understand. On an EATS, the material at high latitude actually 
corresponds to an earlier stage of the ejecta shell, which has a larger 
Lorentz factor and naturally emits harder photons. Additionally, electrons 
enclosed in an EATS is fewer in number than those in the corresponding 
sphere. This is the reason of the reduction of $S_{\nu,{\rm max}}$ 
as mentioned above. 

Finally, we also note that the EATS effect is less 
significant at very high frequency. For example, there is little difference
between the solid line and its corresponding dashed line when $\nu \geq 
10^{17}$ Hz. This can also be easily understood. High energy photons are 
mostly emitted by high speed materials, which mainly reside at the top 
point of the EATS and whose emission is restricted within a small solid 
angle due to relativistic beaming effect. In other words, high energy photons 
are emitted from a small portion of the EATS which is at the top point 
and which differs from a sphere marginally. 

Figure 2 shows the EATS effect on the R-band afterglow light curve. An 
obvious feature is that the EATS effect postpones the peak time ($t_{\rm peak}$) 
of optical afterglow by a factor of $\sim 2$. Also, before the peak time, 
the EATS effect makes the afterglow dimmer, but after the peak time, it 
makes the afterglow slightly brighter. However, the EATS effect does not 
alter the slopes of the light curve, either before or after the peak time. 
Figure~3 illustrates the EATS effect on X-ray afterglows. While the 
basic features of Figure~3 are generally similar to those of Figure~2, an 
obvious difference is that the dimming and brightening of X-ray emission 
before and after the peak time due to the EATS effect are much shallower. It is 
consistent with the spectral characteristics revealed in Figure~1.

\subsection{Jets}

The EATS effect on the optical afterglow of jets are illustrated in 
Figure~4. Generally speaking, the role played by the EATS on jets is 
quite similar to that on isotropic fireballs, i.e., postponing the 
peak time, reducing the brightness before $t_{\rm peak}$, and enhancing 
it after $t_{\rm peak}$. 

Figure 5 shows the afterglow light curves when the observer is off-axis. 
An obvious feature can be immediately noted in this figure that the 
dashed line is much higher above the solid line when $t < t_{\rm peak}$. 
This behavior is not completely unexpected. As we know, when an observer 
is off-axis, the observed flux will be very low due to relativistic 
beaming. When EATS is taken into account, high latitude photons actually 
come from material with larger Lorentz factors, which means the beaming 
effect is more serious. The effect is especially notable at early stages 
($t < t_{\rm peak}$), when the decrease of the Lorentz factor of the jet 
is extremely rapid.

\subsection{Density Jump Cases}

When the GRB ejecta encounters a sudden density increase in the circumburst
medium, a rebrightening of the afterglow will be observed (Lazzati et al. 
2002; Nakar \& Piran 2003; Dai \& Wu 2003; Tam et al. 2005). 
It would be of interest to 
investigate how the EATS takes effect when such a brightness variation is 
involved. Here we assume that the number density of the circumburst medium
jumps suddenly from 1 cm$^{-3}$ to 100 cm$^{-3}$ at the observer's time 
$2\times 10^4$ s (corresponding to a radius of $R_{\rm J} \sim 4.5 \times 
10^{17}$ cm). The numerical results are presented in Figures~6 --- 7.

Figure 6 shows a few surfaces of equal arrival times, comparing them 
directly with the spherical geometry of the jet. At early stages, when
the jet is still highly ultra-relativistic, the EATSes are very flat and 
deviate from spherical surfaces seriously. However, it is interesting to note that 
the foreland of the EATS becames obtuse when $R > R_{\rm J}$. This is 
because the jet decelerates rapidly after the density jump, making the
relativistic effect less significant. At the time
of $10^6$ s, when the Lorentz factor of the jet is $\gamma \sim 
1.2$, the EATS no longer deviates from the sphere so markedly. In our 
calculations, when the blastwave reaches the density jump radius, its 
Lorentz factor is $\gamma \approx 9$. Thus the EATS will completely pass 
through the density jump surface at a time of $R_{\rm J}/\gamma^2 c
\sim 1.9 \times 10^5$ s. This can also be clearly seen in Figure~6. 

Figure 7 shows the R-band afterglow light curves. 
The dashed line in Figure 7a corresponds to the instance when the EATS 
effect is not considered. We see that a sharp rebrightening does appear 
at the density jump moment. However, the flux density decreases steeply 
soon after the density jump. This is mainly due to the rapid deceleration 
of the blastwave in a much denser environment. When the EATS effect is 
included in our calculation, the rapid variation is largely
smeared and the light curve  (the solid line) becomes very different. 
First, the rapid decline of the brightness seen at $t > 2 \times 10^4$ s
in the dashed line is now postponed to $\sim 2 \times 10^5$ s. This is
easy to understand. We know that the EATS is not homogeneous in 
brightness, but shows a ring-like structure, which means emission from 
the high latitude portion plays the dominant role in afterglows (Waxman 
1997; Sari 1998; Panaitescu \& M\'esz\'aros 1998). 
At any time $2 \times 10^4 {\rm s} < t < 
2 \times 10^5$ s, although the central portion of the EATS is in the 
high density region so that the emission is very weak, the high latitude 
portion of the EATS, which dominates the afterglow emission, 
is still in the low density region (see Figure~6) 
and the emissivity remains at a high level. 
So, the afterglow flux will not be affected too much by 
the density increase during this period. However, when $t \geq 2 \times
10^5$ s, the EATS passes through the density jump radius completely and 
the emissivity of the whole EATS becomes very low. The afterglow then 
naturally shows a steep decline. 
Second, the pulse-like rebrightening structure at exactly $2 \times 10^4$ s 
in the dashed line also leaves its fingerprint in the solid line. As a 
result, we can observe a shallow but clear rebrightening in the solid
light curve beginning at the time of the density jump. 

In reality, the density jump is usually due to the existence of a dense 
molecular cloud. Since molecular clouds can be magnetized, it is possible 
that the $\xi_{\rm B}$ parameter may be correspondingly much larger  
after the density jump in some cases. 
In Figure~7b, we assume that at the density jump radius, the 
$\xi_{\rm B}$ parameter increases by a factor of 50 at the same time. This 
induces a prominent pulse-like structure in the dashed light curve when
the EATS effect is expelled. When the EATS effect is included, the 
rebrightening is still very salient. This mechanism may give an 
explanation to the marked rebrightenings observed in some GRB afterglows. 

\subsection{Energy Injection Cases}

Evidence for prolonged activities of the central engines of GRBs has been 
found in a few events (Dai \& Lu 2001; Zhang \& M\'esz\'aros 2002; Bjornsson,
Gudmundsson, \& Johannesson 2004; Fan et al. 2004; Burrows et al. 2005; 
King et al. 2005; Watson et al. 2006). Here we assume that the kinetic energy 
of the GRB remnant increases instantly by a factor three 
at $t = 2 \times 10^4$ s due to a sudden energy injection. The corresponding 
optical light curves are shown in Figure~8. Again we see that the effect of 
the EATS is to smoothen the light curve variability.

\section{Conclusion and Discussion}

In this article we study the EATS effect on GRB afterglows through numerical
calculations. Generally speaking, for isotropic fireballs and jets viewed 
on the axis, the inclusion of the EATS consideration does not change the 
shapes of afterglow spectra and light curves. However, it does slightly harden
the spectra, and postpones the peak time of the light curves. Additionally, 
the EATS effect tends to decrease the flux density when $t < t_{\rm peak}$, 
but increase the brightness when $t > t_{\rm peak}$. In X-ray bands, the 
EATS effect is weaker than that at optical frequencies. 

When the GRB ejecta encounters a sudden density jump in the circumburst 
medium, the emissivity of the blastwave first rises rapidly, but it will
then decrease steeply to a much lower level due to the rapid deceleration 
of the shock in a denser environment. In this case, the EATS effect 
changes the afterglow light curve significantly, shaping the originally 
pulse-like structure into a much weaker but much longer rebrightening. 
In case of energy injection, the EATS has a similar effect, i.e., smoothing
the variability of the light curve. 

Our studies on the EATS effect have important implications on observations. 
A good example is GRB 030329, for which a marked rebrightening was observed 
at $t \sim 1.6$ d in the optical afterglow (Lipkin et al. 2004). Huang, 
Cheng and Gao (2006) have reexamined this event numerically in light of 
three models, i.e., the density-jump model, the two-component jet model, 
and the energy-injection model. EATS effect was considered in their 
calculations. They found that the energy-injection model is the most 
preferred choice for the regrightening. However, even in their best fit
to the optical afterglow by engaging the energy-injection model, the 
theoretical rebrightening is still not rapid enough as compared with 
observations, due to the EATS effect. This hints that we still need to 
seek other physical process for the rebrightening.

In the density jump case considered in our current study, 
there is a possibility that the portion of 
magnetic field energy (i.e., the $\xi_{\rm B}$ parameter) may also 
increase at the jump radius. This may happen when the density jump is
caused by a magnetized molecular cloud. In this case, a prominent 
rebrightening is expected even if the EATS effect is taken into account.
It is characterized by a rapid increase at the beginning and a steep 
decrease after the jump front completely passes through the EATS. 
This mechanism may give a natural explanation to the rebrightenings 
observed in some GRBs. For GRB 030329, it is quite probable that the 
$\xi_{\rm B}$ parameter might also change at the energy-injection moment, 
which may help to fasten the rebrightening. This possibility needs 
further study in the future.

\begin{acknowledgements}
We thank the anonymous referee for helpful comments and suggestions. 
This research was supported by a RGC grant of the Hong Kong Government, 
and also partly supported by the National Natural Science 
Foundation of China (Grants 10625313, 10233010, 10221001, 10573021, and 
10433010), and the Foundation for the Author of National Excellent Doctoral 
Dissertation of P. R. China (Project No: 200125).
\end{acknowledgements}

\begin{figure}
   \begin{center}
   \epsscale{1.0}{1.0}          
   \plotone{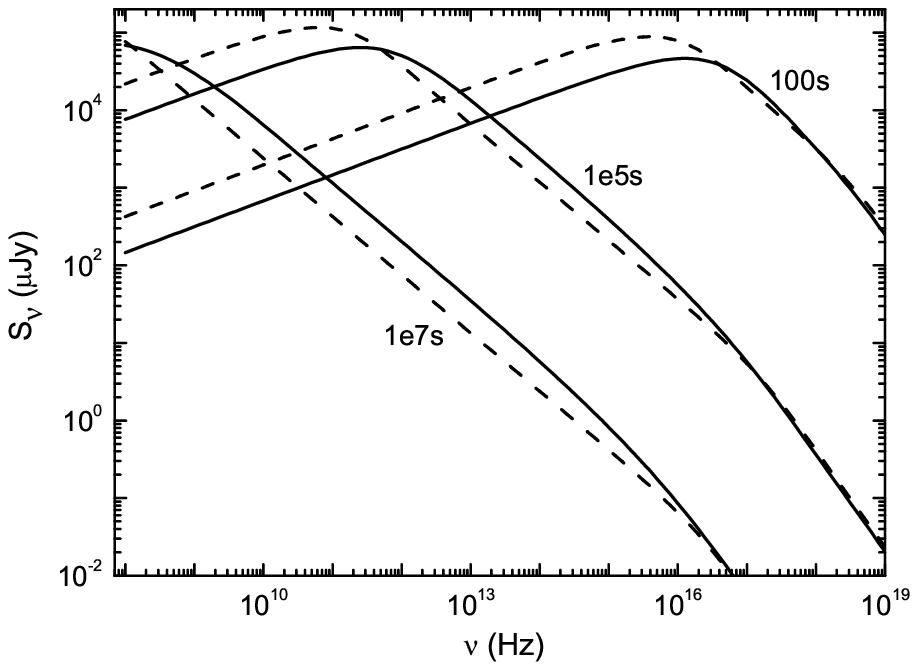}
   \caption{Spectrum evolution of an isotropic fireball with ``standard'' 
parameters. The solid lines are drawn with the EATS effect included. 
As a comparison, the dashed lines do not incorporate the EATS effect. 
The number near each pair of curves indicates the time at which the 
spectra are sampled.  }
   \label{fig1}
   \end{center}
\end{figure}

\begin{figure}
   \begin{center}
   \epsscale{1.0}{1.0}          
   \plotone{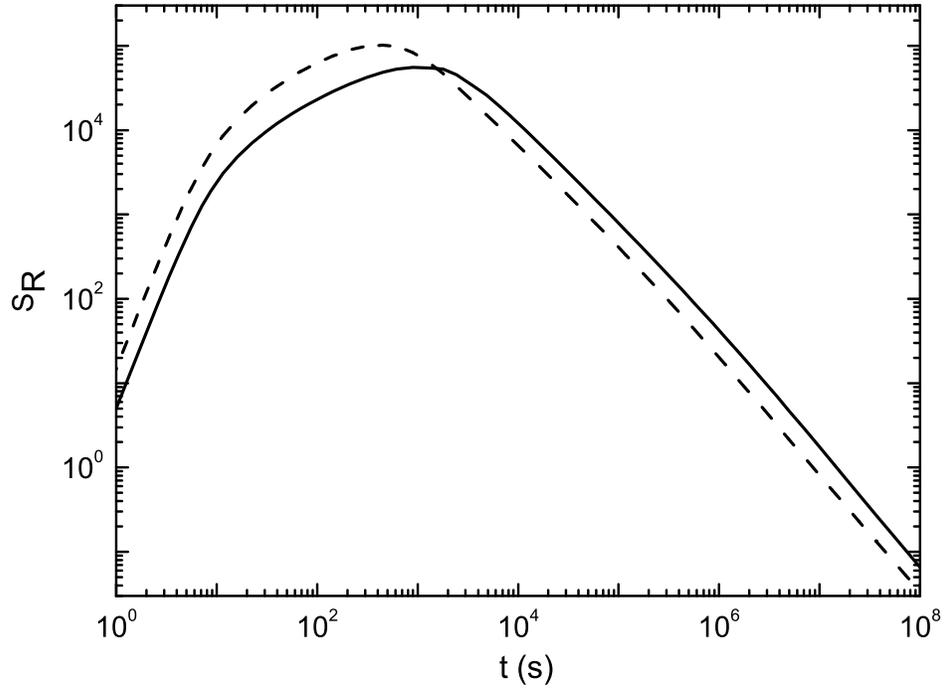}
   \caption{R-band afterglow light curves of an isotropic fireball. 
Line styles and parameters are the same as in Figure~1.  }
   \label{fig2}
   \end{center}
\end{figure}

\begin{figure}
   \begin{center}
   \epsscale{1.0}{1.0}          
   \plotone{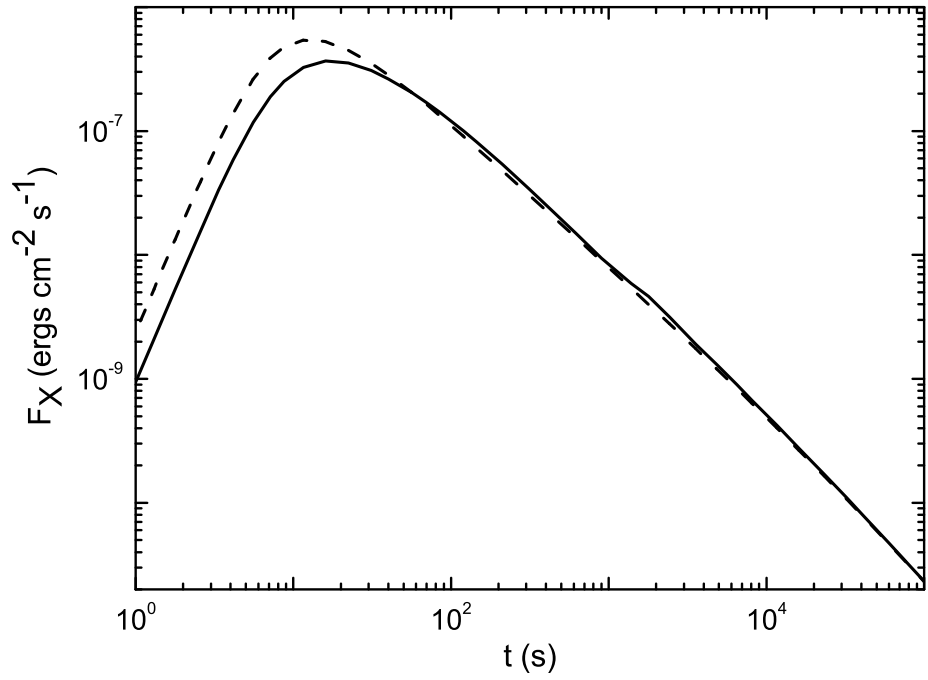}
   \caption{ 0.1 --- 10 keV X-ray afterglow light curves of an isotropic
fireball. Line styles and parameters are the same as in Figure~1. }
   \label{fig3}
   \end{center}
\end{figure}

\begin{figure}
   \begin{center}
   \epsscale{1.0}{1.0}          
   \plotone{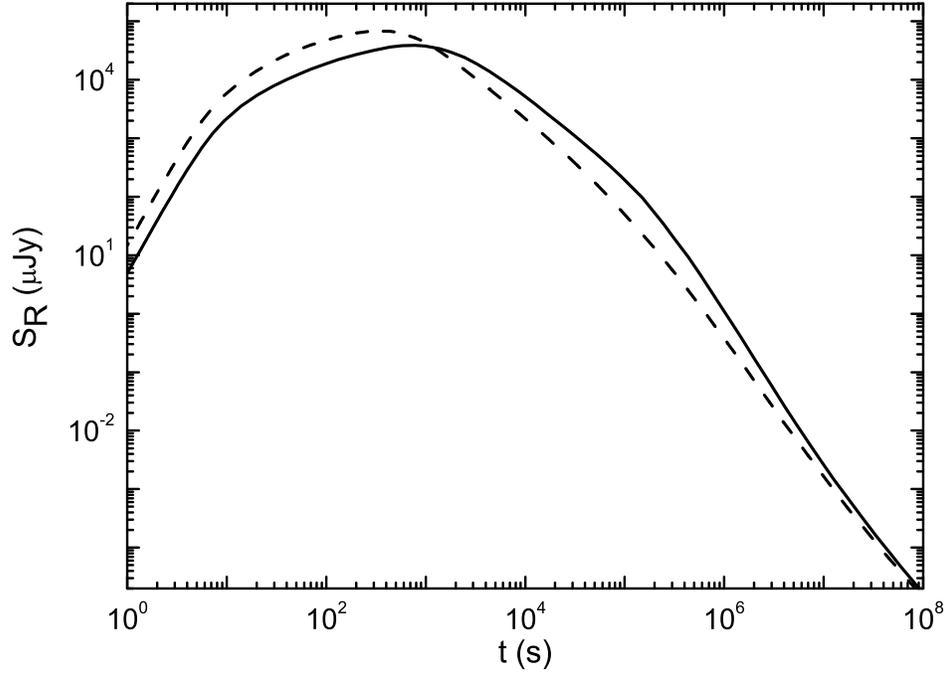}
   \caption{ R-band afterglow light curves of a jet with ``standard'' parameters. 
The solid line is drawn with the EATS effect considered, while the dashed 
line is drawn with the effect excluded. }
   \label{fig4}
   \end{center}
\end{figure}

\begin{figure}
   \begin{center}
   \epsscale{1.0}{1.0}          
   \plotone{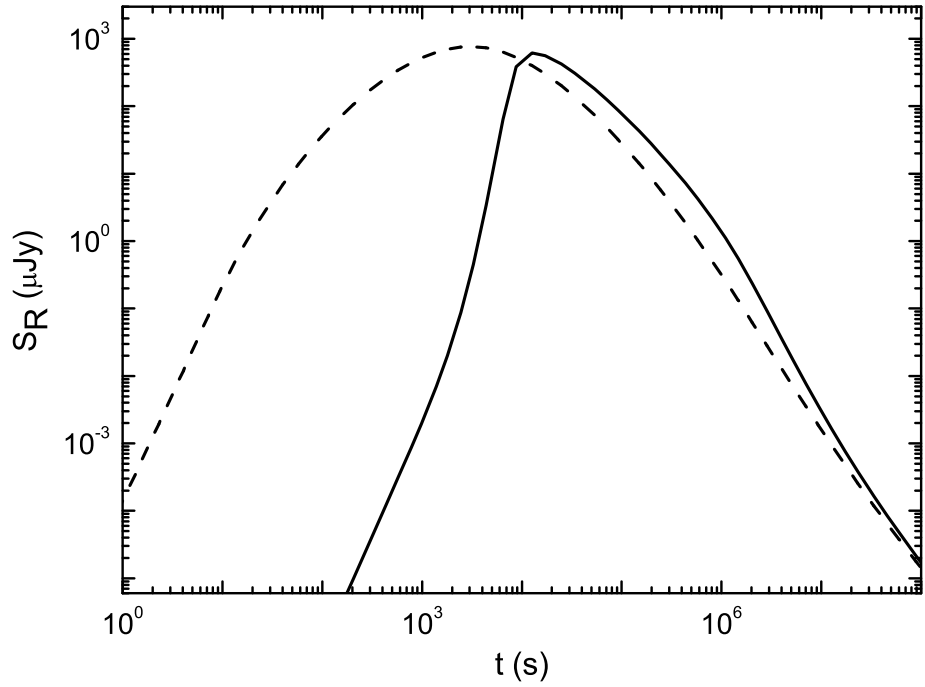}
   \caption{ R-band afterglow light curves of a jet with ``standard'' parameters, 
but viewed at an angle of 0.17. Line styles are the same as in Figure~4. }
   \label{fig5}
   \end{center}
\end{figure}

\begin{figure}
   \begin{center}
   \epsscale{1.0}{1.0}          
   \plotone{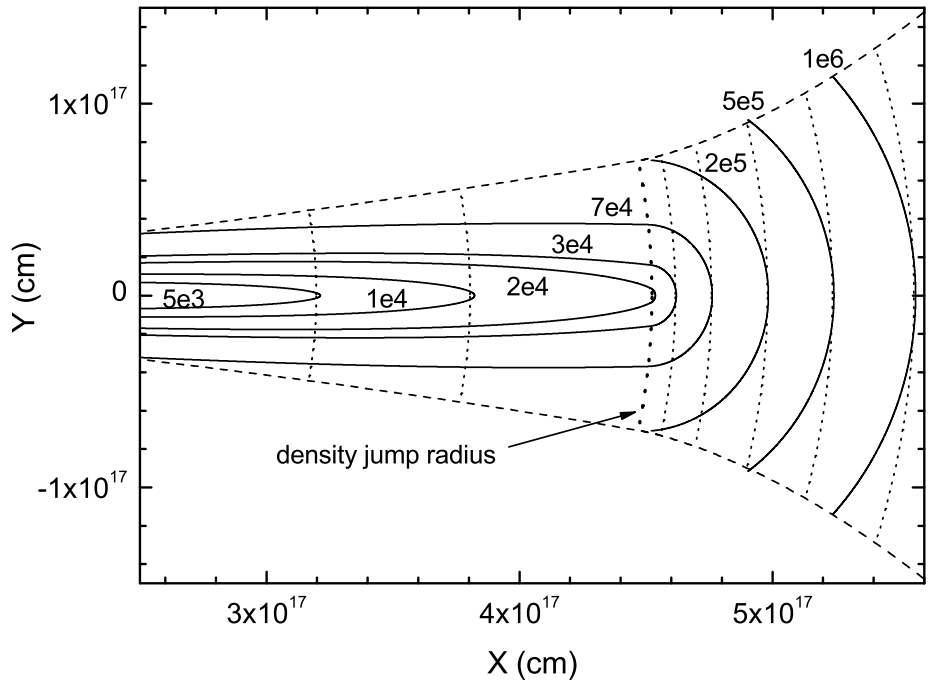}
   \caption{Exemplar surfaces of equal arrival times for a ``standard''
jet encountering a density jump at $t = 2 \times 10^4$ s. The 
amplitude of the density jump is 100 times. X-axis is the direction
of motion of the jet, and Y-axis is the lateral direction. The
solid lines illustrate the equal arrival time surfaces, with 
the time marked in units of s. The dotted lines shows the corresponding
spherical surfaces. The dashed lines are jet boundaries. }
   \label{fig6}
   \end{center}
\end{figure}

%
   \begin{figure}
   \plottwo{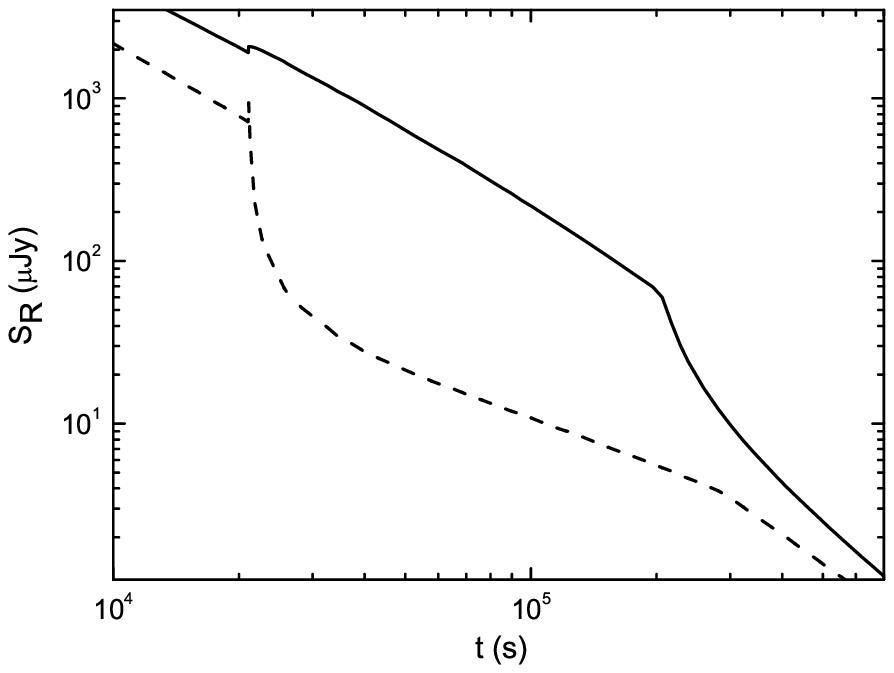} {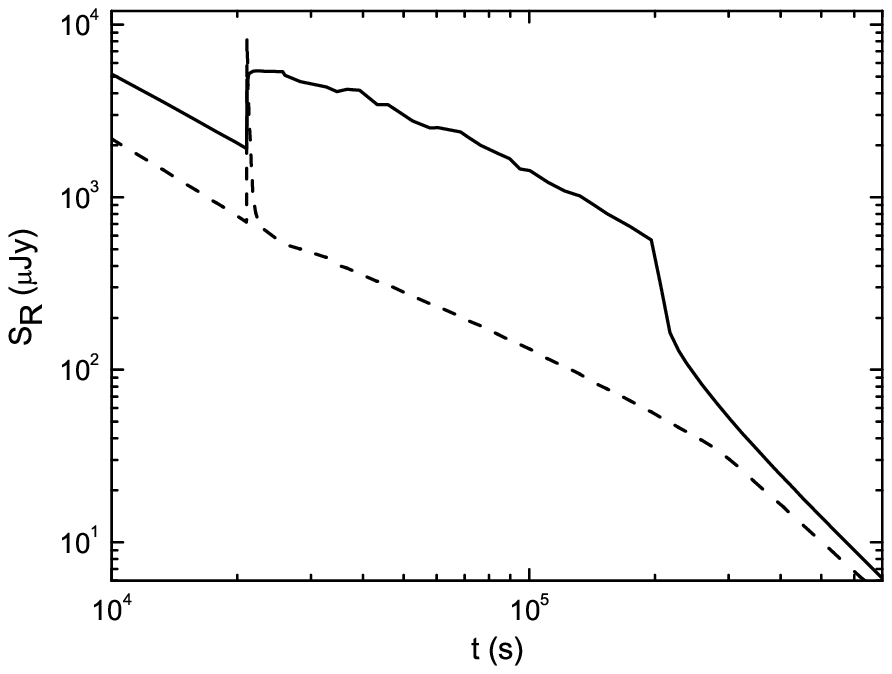}
   \caption{ (a) R-band afterglow light curves of a jet encountering a 
density jump at $t = 2 \times 10^4$ s. The amplitude of the density 
jump is 100 times. Other parameters involved are the same as in 
Figure~4. The solid line and the dashed line correspond to the 
instances when the EATS effect is and is not included, respectively.  
(b) Same as (a), except that the $\xi_{\rm B}$ parameter increases 
by a factor of 50 simultaneously at the jump radius. }
   \label{fig7}
\end{figure}

\begin{figure}
   \begin{center}
   \epsscale{1.0}{1.0}          
   \plotone{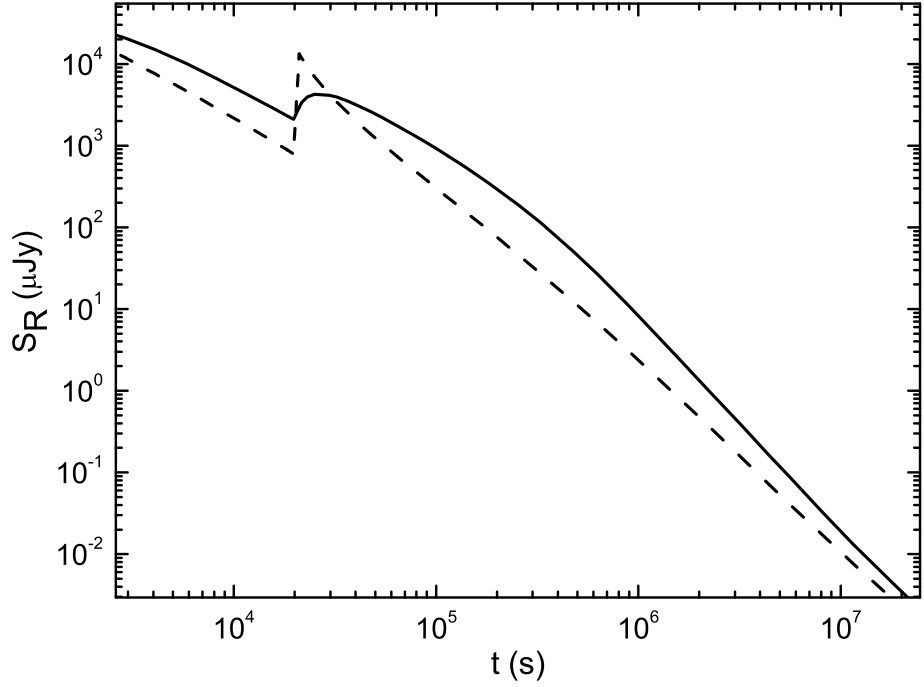}
   \caption{R-band afterglow light curves of a ``standard'' jet in case 
of an energy injection occurring at $2 \times 10^4$ s. The energy 
supply is assumed to be completed instantly, which increases the 
total kinetic energy of the GRB remnant by a factor of 3. The solid 
line is drawn by including the EATS effect, and the dashed line is 
drawn with the effect excluded. }
   \label{fig8}
   \end{center}
\end{figure}

\label{lastpage}

\end{document}